\documentclass[reprint , prl,aps,twocolumn,superscriptaddress,showpacs,preprintnumbers,
               amsmath,amssymb,floatfix]{revtex4-1}

\def\gtwid{\mathrel{\raise.3ex\hbox{$>$\kern-.75em\lower1ex\hbox{$\sim$}}}}
\def\ltwid{\mathrel{\raise.3ex\hbox{$<$\kern-.75em\lower1ex\hbox{$\sim$}}}}

\usepackage{graphicx} 
\usepackage{dcolumn}  
\usepackage{bm}       

\begin{document}

\title{Search for inelastic dark matter with the CDMS~II experiment}


\affiliation{Division of Physics, Mathematics \& Astronomy, California Institute of Technology, Pasadena, California 91125, USA} 
\affiliation{Department of Physics, Case Western Reserve University, Cleveland, Ohio  44106, USA}
\affiliation{Fermi National Accelerator Laboratory, Batavia, Illinois 60510, USA}
\affiliation{Lawrence Berkeley National Laboratory, Berkeley, California 94720, USA}
\affiliation{Department of Physics, Massachusetts Institute of Technology, Cambridge, Massachusetts 02139, USA}
\affiliation{Department of Physics, Queen's University, Kingston, ON, Canada, K7L 3N6}
\affiliation{SLAC National Accelerator Laboratory/KIPAC, Menlo Park, California 94025, USA}
\affiliation{Department of Physics, St.\,Olaf College, Northfield, Minnesota 55057 USA}
\affiliation{Department of Physics, Santa Clara University, Santa Clara, California 95053, USA}
\affiliation{Department of Physics, Southern Methodist University, Dallas, Texas 75275, USA}
\affiliation{Department of Physics, Stanford University, Stanford, California 94305, USA}
\affiliation{Department of Physics, Syracuse University, Syracuse, New York 13244, USA}
\affiliation{Department of Physics, Texas A \& M University, College Station, Texas 77843, USA}
\affiliation{Department of Physics, University of California, Berkeley, California 94720, USA}
\affiliation{Department of Physics, University of California, Santa Barbara, California 93106, USA}
\affiliation{Departments of Phys. \& Elec. Engr., University of Colorado Denver, Denver, Colorado 80217, USA}
\affiliation{Department of Physics, University of Florida, Gainesville, Florida 32611, USA}
\affiliation{School of Physics \& Astronomy, University of Minnesota, Minneapolis, Minnesota 55455, USA}
\affiliation{Physics Institute, University of Z\"{u}rich, Winterthurerstr. 190, CH-8057, Switzerland}

\author{Z.~Ahmed} \affiliation{Division of Physics, Mathematics \& Astronomy, California Institute of Technology, Pasadena, California 91125, USA} 
\author{D.S.~Akerib} \affiliation{Department of Physics, Case Western Reserve University, Cleveland, Ohio  44106, USA} 
\author{S.~Arrenberg} \email{Corresponding author: arrenberg@physik.uzh.ch} \affiliation{Physics Institute, University of Z\"{u}rich, Winterthurerstr. 190, CH-8057, Switzerland}
\author{C.N.~Bailey} \affiliation{Department of Physics, Case Western Reserve University, Cleveland, Ohio  44106, USA} 
\author{D.~Balakishiyeva} \affiliation{Department of Physics, University of Florida, Gainesville, Florida 32611, USA} 
\author{L.~Baudis} \affiliation{Physics Institute, University of Z\"{u}rich, Winterthurerstr. 190, CH-8057, Switzerland}
\author{D.A.~Bauer} \affiliation{Fermi National Accelerator Laboratory, Batavia, Illinois 60510, USA} 
\author{P.L.~Brink} \affiliation{SLAC National Accelerator Laboratory/KIPAC, Menlo Park, California 94025, USA}
\author{T.~Bruch} \affiliation{Physics Institute, University of Z\"{u}rich, Winterthurerstr. 190, CH-8057, Switzerland}
\author{R.~Bunker} \affiliation{Department of Physics, University of California, Santa Barbara, California 93106, USA} 
\author{B.~Cabrera} \affiliation{Department of Physics, Stanford University, Stanford, California 94305, USA} 
\author{D.O.~Caldwell} \affiliation{Department of Physics, University of California, Santa Barbara, California 93106, USA} 
\author{J.~Cooley} \affiliation{Department of Physics, Southern Methodist University, Dallas, Texas 75275, USA} 
\author{E.~do~Couto~e~Silva} \affiliation{SLAC National Accelerator Laboratory/KIPAC, Menlo Park, California 94025, USA} 
\author{P.~Cushman} \affiliation{School of Physics \& Astronomy, University of Minnesota, Minneapolis, Minnesota 55455, USA} 
\author{M.~Daal} \affiliation{Department of Physics, University of California, Berkeley, California 94720, USA} 
\author{F.~DeJongh} \affiliation{Fermi National Accelerator Laboratory, Batavia, Illinois 60510, USA} 
\author{P.~Di~Stefano} \affiliation{Department of Physics, Queen's University, Kingston, ON, Canada, K7L 3N6}
\author{M.R.~Dragowsky} \affiliation{Department of Physics, Case Western Reserve University, Cleveland, Ohio  44106, USA} 
\author{L.~Duong} \affiliation{School of Physics \& Astronomy, University of Minnesota, Minneapolis, Minnesota 55455, USA} 
\author{S. Fallows}\affiliation{School of Physics \& Astronomy, University of Minnesota, Minneapolis, Minnesota 55455, USA} 
\author{E.~Figueroa-Feliciano} \affiliation{Department of Physics, Massachusetts Institute of Technology, Cambridge, Massachusetts 02139, USA} 
\author{J.~Filippini} \affiliation{Division of Physics, Mathematics \& Astronomy, California Institute of Technology, Pasadena, California 91125, USA} 
\author{J.~Fox} \affiliation{Department of Physics, Queen's University, Kingston, ON, Canada, K7L 3N6}
\author{M.~Fritts} \affiliation{School of Physics \& Astronomy, University of Minnesota, Minneapolis, Minnesota 55455, USA} 
\author{S.R.~Golwala} \affiliation{Division of Physics, Mathematics \& Astronomy, California Institute of Technology, Pasadena, California 91125, USA} 
\author{J.~Hall} \affiliation{Fermi National Accelerator Laboratory, Batavia, Illinois 60510, USA} 
\author{R.~Hennings-Yeomans} \affiliation{Department of Physics, Case Western Reserve University, Cleveland, Ohio  44106, USA} 
\author{S.A.~Hertel} \affiliation{Department of Physics, Massachusetts Institute of Technology, Cambridge, Massachusetts 02139, USA} 
\author{D.~Holmgren} \affiliation{Fermi National Accelerator Laboratory, Batavia, Illinois 60510, USA} 
\author{L.~Hsu} \affiliation{Fermi National Accelerator Laboratory, Batavia, Illinois 60510, USA} 
\author{M.E.~Huber} \affiliation{Departments of Phys. \& Elec. Engr., University of Colorado Denver, Denver, Colorado 80217, USA}
\author{O.~Kamaev}\affiliation{School of Physics \& Astronomy, University of Minnesota, Minneapolis, Minnesota 55455, USA} 
\author{M.~Kiveni} \affiliation{Department of Physics, Syracuse University, Syracuse, New York 13244, USA} 
\author{M.~Kos} \affiliation{Department of Physics, Syracuse University, Syracuse, New York 13244, USA} 
\author{S.W.~Leman} \affiliation{Department of Physics, Massachusetts Institute of Technology, Cambridge, Massachusetts 02139, USA} 
\author{S.~Liu} \affiliation{Department of Physics, Queen's University, Kingston, ON, Canada, K7L 3N6}
\author{R.~Mahapatra} \affiliation{Department of Physics, Texas A \& M University, College Station, Texas 77843, USA} 
\author{V.~Mandic} \affiliation{School of Physics \& Astronomy, University of Minnesota, Minneapolis, Minnesota 55455, USA} 
\author{K.A.~McCarthy} \affiliation{Department of Physics, Massachusetts Institute of Technology, Cambridge, Massachusetts 02139, USA} 
\author{N.~Mirabolfathi} \affiliation{Department of Physics, University of California, Berkeley, California 94720, USA} 
\author{D.~Moore}  \affiliation{Division of Physics, Mathematics \& Astronomy, California Institute of Technology, Pasadena, California 91125, USA}
\author{H.~Nelson} \affiliation{Department of Physics, University of California, Santa Barbara, California 93106, USA} 
\author{R.W.~Ogburn}\affiliation{Department of Physics, Stanford University, Stanford, California 94305, USA} 
\author{A.~Phipps}\affiliation{Department of Physics, University of California, Berkeley, California 94720, USA} 
\author{M.~Pyle} \affiliation{Department of Physics, Stanford University, Stanford, California 94305, USA} 
\author{X.~Qiu} \affiliation{School of Physics \& Astronomy, University of Minnesota, Minneapolis, Minnesota 55455, USA} 
\author{E.~Ramberg} \affiliation{Fermi National Accelerator Laboratory, Batavia, Illinois 60510, USA} 
\author{W.~Rau} \affiliation{Department of Physics, Queen's University, Kingston, ON, Canada, K7L 3N6}
\author{M.~Razeti} \affiliation{Department of Physics, Stanford University, Stanford, California 94305, USA} 
\author{A.~Reisetter} \affiliation{School of Physics \& Astronomy, University of Minnesota, Minneapolis, Minnesota 55455, USA} \affiliation{Department of Physics, St.\,Olaf College, Northfield, Minnesota 55057 USA}
\author{R.~Resch} \affiliation{SLAC National Accelerator Laboratory/KIPAC, Menlo Park, California 94025, USA}  
\author{T.~Saab} \affiliation{Department of Physics, University of Florida, Gainesville, Florida 32611, USA}
\author{B.~Sadoulet} \affiliation{Lawrence Berkeley National Laboratory, Berkeley, California 94720, USA} \affiliation{Department of Physics, University of California, Berkeley, California 94720, USA}
\author{J.~Sander} \affiliation{Department of Physics, University of California, Santa Barbara, California 93106, USA} 
\author{R.W.~Schnee} \affiliation{Department of Physics, Syracuse University, Syracuse, New York 13244, USA} 
\author{D.N.~Seitz} \affiliation{Department of Physics, University of California, Berkeley, California 94720, USA} 
\author{B.~Serfass} \affiliation{Department of Physics, University of California, Berkeley, California 94720, USA} 
\author{K.M.~Sundqvist} \affiliation{Department of Physics, University of California, Berkeley, California 94720, USA} 
\author{M.~Tarka}\affiliation{Physics Institute, University of Z\"{u}rich, Winterthurerstr. 190, CH-8057, Switzerland}
\author{P.~Wikus} \affiliation{Department of Physics, Massachusetts Institute of Technology, Cambridge, Massachusetts 02139, USA} 
\author{S.~Yellin} \affiliation{Department of Physics, Stanford University, Stanford, California 94305, USA} \affiliation{Department of Physics, University of California, Santa Barbara, California 93106, USA}
\author{J.~Yoo} \affiliation{Fermi National Accelerator Laboratory, Batavia, Illinois 60510, USA} 
\author{B.A.~Young} \affiliation{Department of Physics, Santa Clara University, Santa Clara, California 95053, USA} 
\author{J.~Zhang}\affiliation{School of Physics \& Astronomy, University of Minnesota, Minneapolis, Minnesota 55455, USA}

\collaboration{CDMS Collaboration}
\noaffiliation

\begin{abstract}
Results are presented from a reanalysis of the entire five-tower data set acquired with the Cryogenic Dark Matter Search (CDMS~II) experiment at the Soudan Underground Laboratory, with an exposure of 969\,kg-days. The analysis window was extended to a recoil energy of 150\,keV, and an improved surface-event background-rejection cut was defined to increase the sensitivity of the experiment to the inelastic dark matter (iDM) model. Three dark matter candidates were found between 25\,keV and 150\,keV. The probability to observe three or more background events in this energy range is 11\%. Because of the occurrence of these events the constraints on the iDM parameter space are slightly less stringent than those from our previous analysis, which used an energy window of 10{\textendash}100\,keV.
\end{abstract}
\maketitle

Cosmological observations \cite{wmap} strongly suggest that nonluminous, nonbaryonic matter constitutes most of the matter in the Universe. This dark matter should be distributed in dark halos of galaxies such as the Milky Way, enabling the direct detection of the dark matter particles via their interactions in terrestrial detectors \cite{goodmanwitten}. The movement of the Earth around the Sun would provide an annual modulation of the counting rate, caused by the change in the relative velocity of the dark matter particles with respect to the earthbound target \cite{drukier86}.

The DAMA collaboration claims the observation of such a modulation in two different NaI(Tl) scintillation detector arrays \cite{dama2000,damalibra2008}. The observed signal is in the \mbox{2{\textendash}6\,keV} electron-equivalent energy range with a periodicity of 0.999$\pm$0.002 years and a phase of 146$\pm$7 days \cite{damalibra2010}. The observed modulation signature is consistent with the expected signature of galactic dark matter particles interacting in a terrestrial detector. 
Other experimental results \cite{cdms2010, xenon100,xenon10sd,coupp,kims,zeplinIII}, however, are inconsistent with the interpretation of the DAMA result as a signal from weakly interacting massive particles (WIMPs) \cite{lee, weinberg,jungman,bertone} elastically scattering off nuclei.

Inelastic dark matter (iDM) scattering has been proposed as a way to resolve this tension \cite{idmscenario}. The inelastic scenario assumes that WIMPs ($\chi$) can only scatter off baryonic matter ($N$) by transition into an excited state at a certain energy above the ground state ($\chi \; N \to \chi^* \; N$), while elastic scattering is forbidden or highly suppressed. There is a minimal velocity required to produce recoil energy $E_R$ in such an inelastic scatter,
\begin{equation}
v_{\mathrm{min}} = \frac{1}{\sqrt{2m_N E_R}} \bigg(\frac{m_N E_R}{\mu}+\delta\bigg) \, ,
\label{vmin}
\end{equation}
where $m_N$ is the mass of the target nucleus, $\mu$ is the reduced mass of the WIMP-nucleus system, and $\delta$ is the WIMP-mass splitting; $\delta = 0 \; \mathrm{keV}$ is equivalent to elastic scattering. If $E_R$ is too small or too large, $v_{\mathrm{min}}$ is above the cutoff imposed by the galactic escape velocity, and the event cannot occur. Important consequences of this model for direct detection experiments are differential rates that peak at tens of keV recoil energy, and a significant suppression of the recoil spectrum at low recoil energies. In addition, the annual modulation signature is significantly enhanced, because of the increased dependence on the high-velocity tail of the WIMP-velocity distribution, which in turn is due to the larger minimal velocity (see Eq.~\eqref{vmin}). Therefore, the iDM scenario is also particularly sensitive to the escape-velocity cutoff in the WIMP-velocity distribution. Finally, it is important to note that the scattering rate is enhanced for heavy target nuclei (\mbox{e.g.} Xe and I).

The Cryogenic Dark Matter Search (CDMS~II) experiment operated in the Soudan Underground Laboratory a total of 19\,Ge ($\sim$230\,g each) and 11\,Si ($\sim$105\,g each) detectors at a temperature of $\sim$40\,mK~\cite{zips, prd118}. These semiconductors were stacked into five towers (T1{\textendash}T5) with six detectors (Z1{\textendash}Z6) each. They were instrumented with four channels of superconducting transition-edge sensors on the top side to detect phonons and two concentric electrodes on the bottom side to detect ionization. The primary ionization signal was read out by an inner electrode covering $\sim$85\% of the detector surface. A thin outer electrode served as a guard ring to identify and reject events at the edge of the detector, which was subject to higher background and reduced charge collection. The recoil energy was reconstructed from the phonon and the ionization signal \cite{negluke}. The ratio of ionization to recoil energy (``ionization yield'') was lower for nuclear recoils, produced by WIMP candidates, than for electron recoils, caused mostly by background photons. Fewer than $10^{-4}$ of the electron recoils in the bulk of the detector were misidentified as nuclear recoils. The main source of misidentified electron recoils were events with interactions in the first few $\mu$m of the detector surfaces. Because of incomplete charge collection these events had reduced ionization yield, and occasionally the reduction was severe enough to mimic a WIMP-nucleus interaction. The phonon signals of these surface electron-recoil events had faster-rising pulses than bulk nuclear recoils and occurred closer in time to the more prompt ionization pulses. As discussed in detail in this paper a cut based on these timing parameters was employed to reject interactions at the detectors' surfaces. Misidentified surface events constituted the dominant background for the CDMS~II experiment, while the neutron background from cosmogenics and radioactive processes was much less significant.

\begin{figure}[b!]
\includegraphics[width=0.48\textwidth]{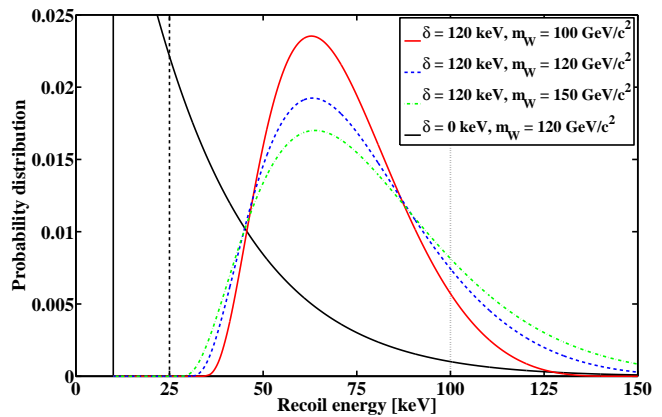}
\caption{(color online). Differential recoil spectra in a Ge target for a WIMP-mass splitting $\delta$ of 120\,keV and a few representative WIMP masses $m_W$. For comparison the spectrum for a WIMP with a mass of 120\,GeV/c$^2$ assuming elastic scattering ($\delta$ = 0 keV) is also shown (black/solid). The spectra are normalized to unity in the 10{\textendash}150\,keV recoil-energy range. The vertical lines denote the analysis threshold at 10\,keV, the lower boundary for the setting of the surface-event rejection cut at 25\,keV, and the upper analysis limit from our previous analysis at 100\,keV \cite{cdms2010}. See text for details.}
\label{spectrum}
\end{figure}

Initial constraints from CDMS on the iDM model interpretation of the DAMA claim were set using a recoil-energy range of 10{\textendash}100\,keV \cite{cdms2010}. This paper presents a dedicated iDM analysis of the entire CDMS~II five-tower data set, taken during two periods of stable operation between October 2006 and July 2007 (internally denoted as runs 123{\textendash}124) \cite{cdms2008}, and four periods between July 2007 and September 2008 (internally denoted as runs 125{\textendash}128) \cite{cdms2010}. Note that the constraints on the WIMP-parameter space shown in \cite{cdms2010} were a combination of the final results from all data sets taken at the Soudan Underground Laboratory, which, however, were analyzed separately. In particular, the surface-event rejection cuts, as discussed below, were set at fixed backgrounds for runs 123{\textendash}124 and runs 125{\textendash}128 separately. For the analysis presented here, the whole acquired data were combined in advance and surface-event rejection was based on the entire data set. There were two main reasons for performing this reanalysis. The iDM parameter space allowed by our previous analysis (see Fig.~4 of \cite{cdms2010}) includes WIMP masses $m_W\sim100\,\mathrm{GeV/c}^2$ and mass splittings $\delta\sim120\,\mathrm{keV}$. As shown in Fig.~\ref{spectrum}, these parameters result in a significant expected rate above our previous analysis upper limit of 100\,keV, so a simple extension to 150\,keV increases the expected sensitivity. Moreover, the expected rate drops to zero for low recoil energies, in contrast to the elastic-scattering case, obviating the need for a low threshold. Since most of the dominant surface-event background occurred at energies just above our 10\,keV threshold \cite{cdms2008}, where no iDM signal is expected, the sensitivity could be further improved by redefining a looser surface-event rejection cut based upon the estimated background with recoil energy between 25\,keV and 150\,keV, while leaving the lower boundary for the analysis at 10\,keV. Thus, a significant number of surface-background events was expected in the 10{\textendash}25\,keV range, which, however, had only a minor effect on the results in the parameter-space region of interest ($m_W\sim100\,\mathrm{GeV/c}^2$, $\delta\sim120\,\mathrm{keV}$).

The same data-quality selection cuts used in previous analyses \cite{cdms2010, cdms2008} for ensuring detector stability and removing periods of poor detector performance, \mbox{e.g.} due to insufficient neutralization, causing incomplete charge collection owing to impurities in the detector crystal, resulted in a total Ge exposure of 969\,kg-days for this reanalysis. The Si detectors were omitted due to their lower sensitivity to inelastic scattering. Because both data sets had already been analyzed, this analysis was not ``blind''. However, the analysis was performed in a similar manner to minimize bias: selection criteria and background estimates were defined and evaluated using only WIMP-search data outside the signal region and calibration data.

\begin{figure}[t!]
\includegraphics[width=0.48\textwidth]{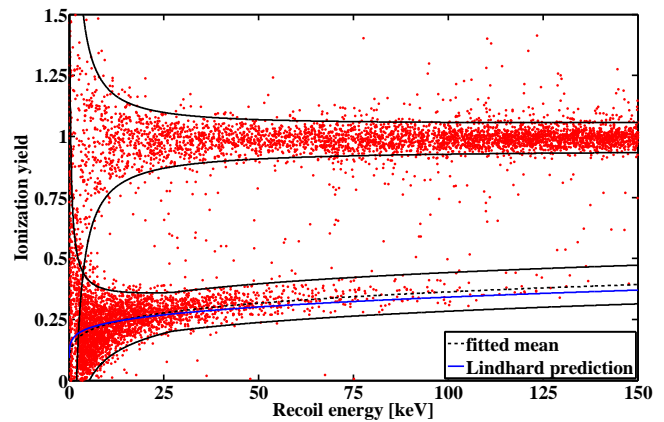}
\caption{(color online). Ionization yield versus recoil energy of $^{252}$Cf calibration data from a representative detector in one of the six data runs. The black/solid lines represent the electron-recoil band around a yield of one and the nuclear-recoil band around 0.3. The black/dashed line denotes the mean of the latter band, while the similar but blue/solid line is the corresponding prediction from Lindhard theory \cite{lindhard, lewinsmith}.}
\label{bands}
\end{figure}

The detectors were exposed to gamma rays from $^{133}$Ba and neutrons from $^{252}$Cf at regular intervals to calibrate their response and define criteria for data-quality cuts and the WIMP-acceptance region. The latter was defined to be the $\pm2\sigma$~band around the mean nuclear-recoil ionization yield in the yield versus recoil-energy plane. An illustration is given in Fig.~\ref{bands}, which shows $^{252}$Cf calibration data from a representative detector in one of the six data runs.

In addition to the quality cuts, most of the selection criteria for WIMP-nucleon interactions remained unchanged from the previous analyses \cite{cdms2010, cdms2008}. This included the single-scatter cut, requiring there to be no signal exceeding the phonon-noise level by more than 4$\sigma$ in any of the other 29 detectors; the ionization-based fiducial-volume cut, rejecting events near the edges of the detectors; and the muon-veto cut, demanding negligible coincident energy deposited in the active muon veto surrounding the apparatus.

Extending the analysis window to 150\,keV was hindered by the fact that statistics from the $^{252}$Cf neutron source were low above $\sim$100\,keV which can be seen in Fig.~\ref{bands}. Thus, we extrapolated the nuclear-recoil bands at higher energies from the fits below 100\,keV. The extrapolation showed good agreement with Lindhard theory \cite{lindhard, lewinsmith} when statistics from all six runs were combined for each detector, and both the band locations and the nuclear-recoil cut efficiencies had only a minor energy dependence above $\sim$25\,keV.

The surface-event rejection was based upon a ``timing parameter'' consisting of the sum of the rise time of the largest phonon pulse and its delay relative to the ionization pulse. This timing cut was set in the 25{\textendash}150\,keV energy range using Ba and Cf calibration data. Since surface events in WIMP-search data did not have the same recoil-energy and ionization-yield distributions as in Ba calibration data \cite{cdms2010}, this cut was not expected to be optimal, although corrections based on WIMP-search multiple scatters were applied to the distributions to diminish these differences. Thus, the cut performance had to be tested on WIMP-search data before ``unblinding''. The cut setting and testing are discussed in more detail in the following two sections.

\begin{figure}[b!]
\includegraphics[width=0.48\textwidth]{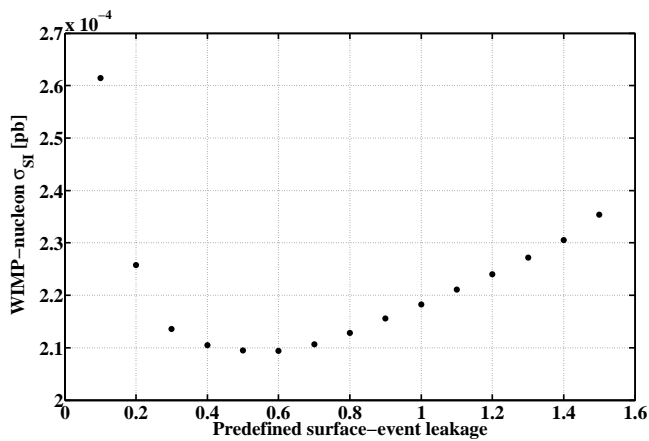}
\caption{Expected sensitivity of this analysis for a WIMP of mass 100\,GeV/c$^2$ and mass splitting 120\,keV for different predefined surface-event leakages at which the timing cut could be fixed. The cut corresponding to the minimum of 0.6 events was chosen as the final cut for this analysis before looking at the WIMP-search signal region.}
\label{sensitivity}
\end{figure}

Each detector had its own timing-parameter cut. We tuned the set of cuts to yield a given expected ``leakage'' (number of background events) for the whole data set, while maximizing the signal for a WIMP of mass 100\,GeV/c$^2$ and a mass splitting of 120\,keV. For each given expected leakage, using values in steps of 0.1 between 0.1 and 1.5,  we ran Monte Carlo simulations to find the average upper limit we could obtain if there were no true WIMP signal. For each expected leakage, $10^5$ surface-event mock data sets were generated, each with number and energies of background events chosen randomly according to the given expected leakage and the expected energy distribution as estimated from WIMP-search multiple scatters. As was to be done with the actual data, a 90\% C.L. upper limit on the spin-independent WIMP-nucleon cross section $\sigma_{\mathrm{SI}}$ was calculated for each mock data set, using the optimum interval method \cite{optint} with the WIMP recoil-energy distribution \cite{lewinsmith, savage} for the chosen WIMP parameters \mbox{($m_W$ = 100\,GeV/c$^2$,} $\delta$ = 120 keV). Figure~\ref{sensitivity} shows the mean upper limit obtained as a function of the expected leakage used in selecting the set of timing-parameter cuts. The timing-parameter cuts were finalized at the values obtained for a fixed expected leakage of 0.6\,events, where the minimum was reached.

\begin{figure}[b!]
\includegraphics[width=0.48\textwidth]{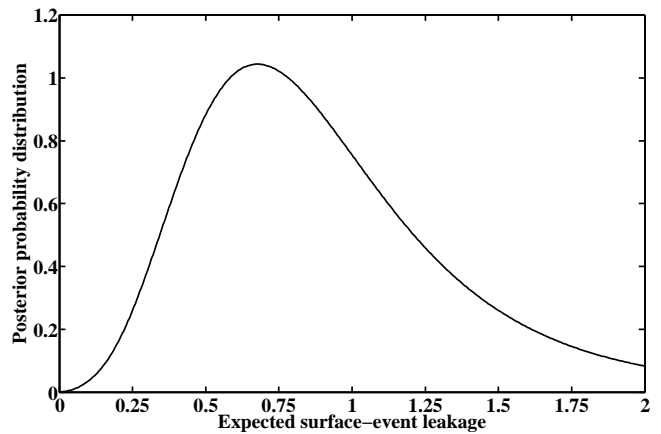}
\caption{Distribution of the surface-event leakage estimate in the 25{\textendash}150\,keV energy range. See text for details.}
\label{leakage}
\end{figure}

As explained above, the leakage value chosen for optimizing cuts was not a sufficiently accurate estimate of the expected background. Thus, as with our earlier \mbox{analysis \cite{cdms2010}}, we used WIMP-search data to improve our estimate of the expected leakage. We estimated the leakage by multiplying the number of WIMP-search nuclear-recoil single scatters failing the timing cut by pass-fail ratios deduced from event samples which were assumed to resemble the population of background events. For detectors that were not located at the top or bottom of their towers (interior detectors), two classes of multiple-scatter events in the WIMP-search data were used independently to estimate the ratios, and therefore the expected background: events with ionization yield within the nuclear-recoil band, and events in which a detector had yield just above or below the nuclear-recoil band (wide-band events). The latter class was defined to include events outside the $\pm2\sigma$~nuclear-recoil band that had an ionization yield above 0.1 and below the minimum of 0.7 and the lower boundary of the $\pm5\sigma$ electron-recoil band at the events' recoil energies. We also included two detectors at the bottom of their towers (end cap detectors) in this analysis. In this case, we treated surface events on the top (internal) and bottom (external) sides of the detectors separately. The pass-fail ratios of the internal sides were estimated from multiple-scatter events with ionization yield within the nuclear-recoil band, and those of the external sides, where tagging of multiple scatters was not possible, from single scatters within the wide-band region. In both cases, interior and end cap detectors, we applied appropriate correction factors to the pass-fail ratios from wide-band events to account for differences in timing performance between surface events within and outside the nuclear-recoil band. For the end cap detectors, additional correction factors were introduced to correct for differences in the single-scatter event fractions on the top and bottom sides. Systematic errors from the estimates of these correction factors, as well as from systematic differences in timing-cut performance between single and multiple scatters, were included in the leakage calculation \cite{bayes}. Because of the low number of events passing the timing cut a dedicated Bayesian surface-event leakage estimate was applied \cite{bayes}, adding another systematic error from the choice of prior distribution. The final background distribution obtained by combining the two estimates from the interior detectors with the estimate from the end cap detectors is shown in Fig.~\ref{leakage}. It contains all statistical and systematic errors. It has a maximum around 0.6 events where the leakage had been fixed for the setting of the cut, while the median, which we use as the final background estimate, is slightly higher but agrees with this value within error bars:
\begin{equation}
\mu_{25-150\, \mathrm{keV}} = 0.8_{-0.3}^{+0.5} (\mathrm{stat})_{-0.2}^{+0.3} (\mathrm{syst}) \;.
\end{equation}
As expected, a similar estimate in the low-energy range from 10{\textendash}25\,keV resulted in a much higher number of expected leakage events:
\begin{equation}
\mu_{10-25\, \mathrm{keV}} = 5.7_{-1.5}^{+2.1} (\mathrm{stat})_{-0.9}^{+1.0} (\mathrm{syst}) \;.
\end{equation}

\begin{figure}[t!]
\includegraphics[width=0.48\textwidth]{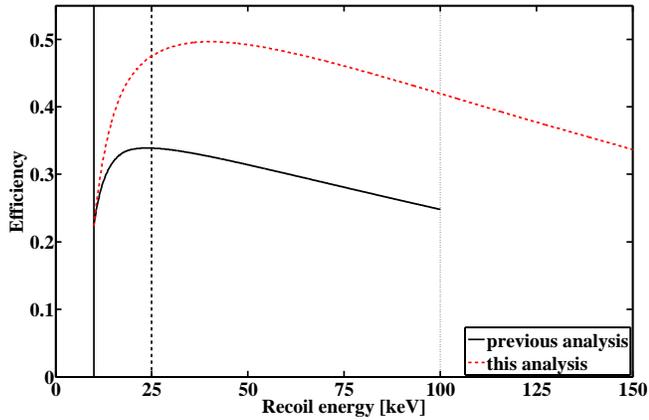}
\caption{(color online). Comparison of the total nuclear-recoil efficiency obtained in this analysis (red/dashed) and from our previous analysis based only on runs 125{\textendash}128 (black/solid) \cite{cdms2010}. The latter is only defined up to 100 keV. Redefining the timing cut achieved an increase in efficiency by a factor of $\sim$1.5. As in Fig.~\ref{spectrum}, the vertical lines denote the analysis threshold at 10\,keV, the lower boundary for the setting of the surface-event rejection cut at 25\,keV, and the upper analysis limit from our previous analysis at 100\,keV.}
\label{efficiency}
\end{figure}

Figure~\ref{efficiency} compares the final efficiency from this analysis, based on runs 123{\textendash}128, with the efficiency from the previous analysis of runs 125{\textendash}128 \cite{cdms2010}. In both analyses the surface-event rejection cuts had roughly the same expected leakage in the energy range the cut was defined on (10{\textendash}100\,keV for the previous analysis and 25{\textendash}150\,keV for the analysis presented here). Even though an exposure which was larger by a factor of 1.6 was considered for the setting of the timing cut,  the final efficiency increased by a factor of $\sim$1.5. This improvement in efficiency was possible because we neglected background at energies below where a signal is expected from iDM.

Neutrons, induced by muons and produced by radioactive processes within the experimental apparatus, constituted an additional background which was indistinguishable from a WIMP interaction in the detectors. Extensive simulations carried out with GEANT4 \cite{geant1, geant2} and FLUKA \cite{fluka1, fluka2} indicated that the neutron background in the 25{\textendash}150\,keV energy range inducted by muons is expected to be $0.04_{-0.03}^{+0.05} (\mathrm{stat.})$, and the background from radioactive processes is estimated to be between 0.03 and 0.06. The background between 10\,keV and 25\,keV is predicted to be $0.06_{-0.04}^{+0.07} (\mathrm{stat.})$ from muon-induced neutrons and between 0.04 and 0.08 from radiogenic neutrons. These estimates are valid for the reanalyzed exposure and include cut acceptances.

\begin{figure}[t!]
\includegraphics[width=0.48\textwidth]{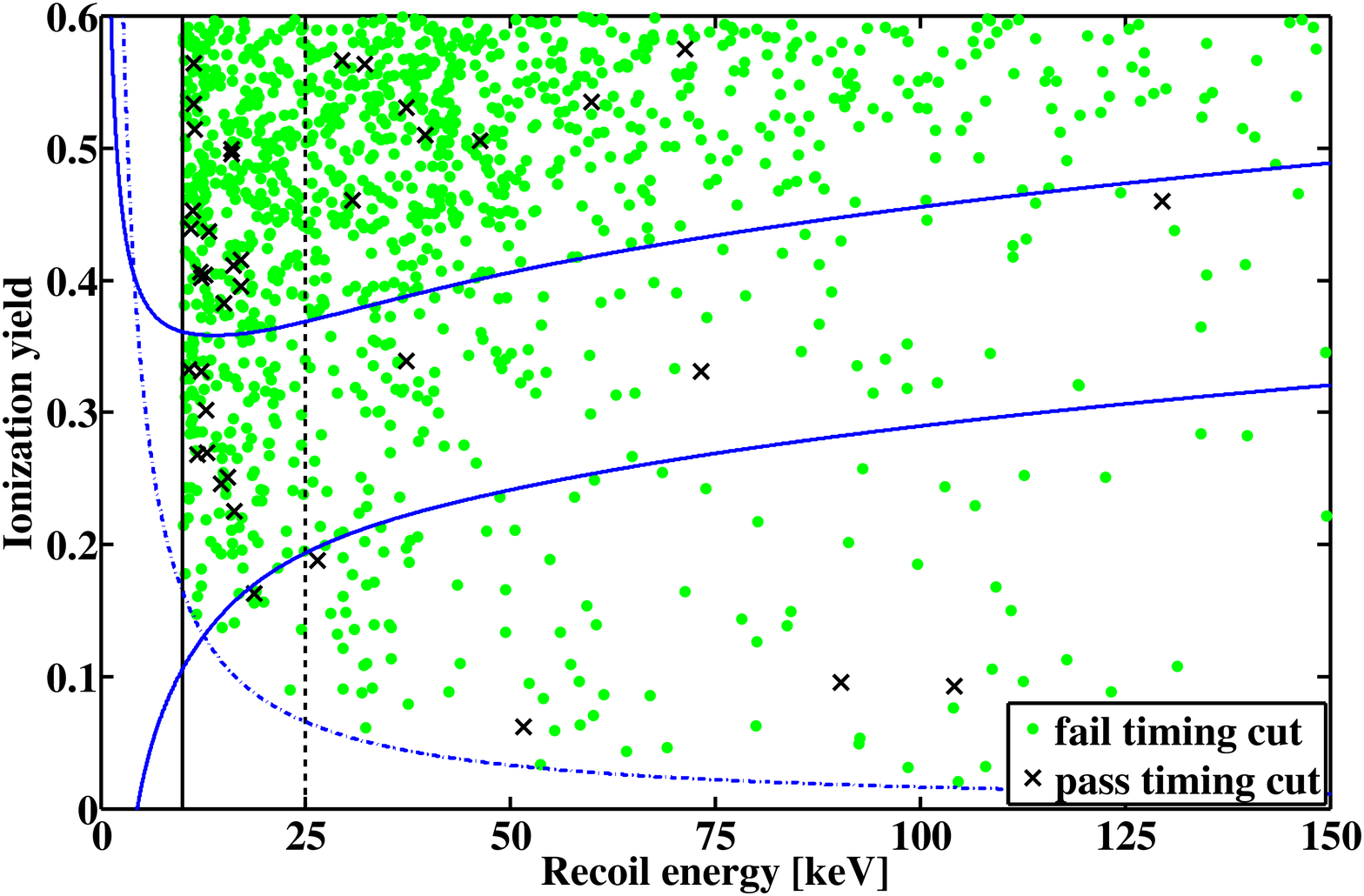}
\caption{(color online). Ionization yield versus recoil energy for all low-yield WIMP-search single scatters from the combined five-tower data set (runs 123{\textendash}128) passing all constraints except for cuts on the ionization yield and timing parameter. Most events fail the timing cut (green dots), while most of the few dozen that pass the timing cut ($\times$) fall outside the nuclear-recoil band (blue/solid lines). Eleven events pass all the selection criteria, with three occurring within the 25{\textendash}150\,keV range upon which the surface-event rejection cut was defined, and eight between the 10\,keV recoil-energy threshold and 25\,keV. The ionization-energy threshold is also shown (blue/dashed-dotted). This threshold and the shown nuclear-recoil band represent the exposure-weighted means over all runs and detectors. }
\label{yieldenergy}
\end{figure}

\begin{figure}[t!]
\includegraphics[width=0.48\textwidth]{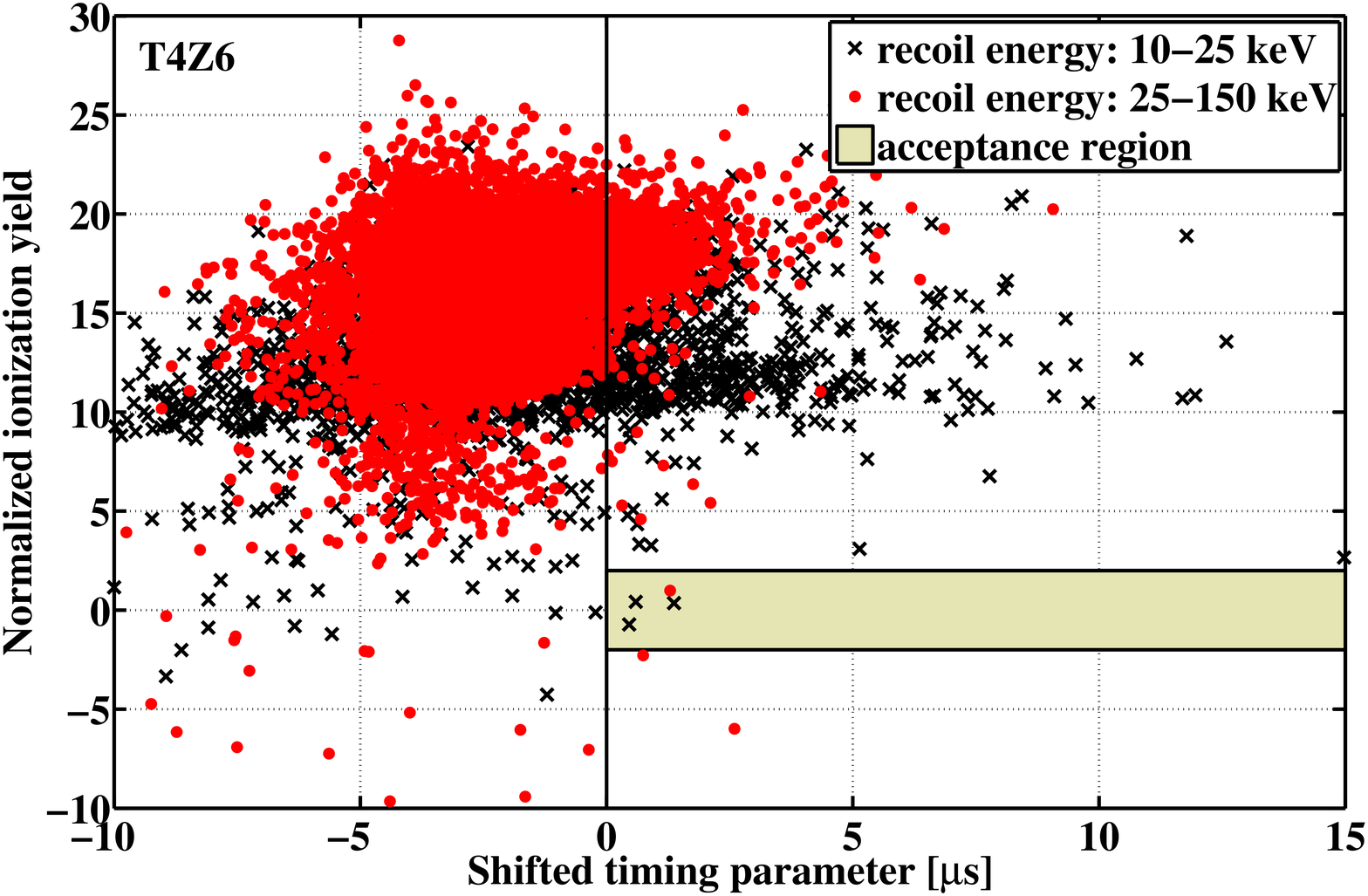}
\includegraphics[width=0.48\textwidth]{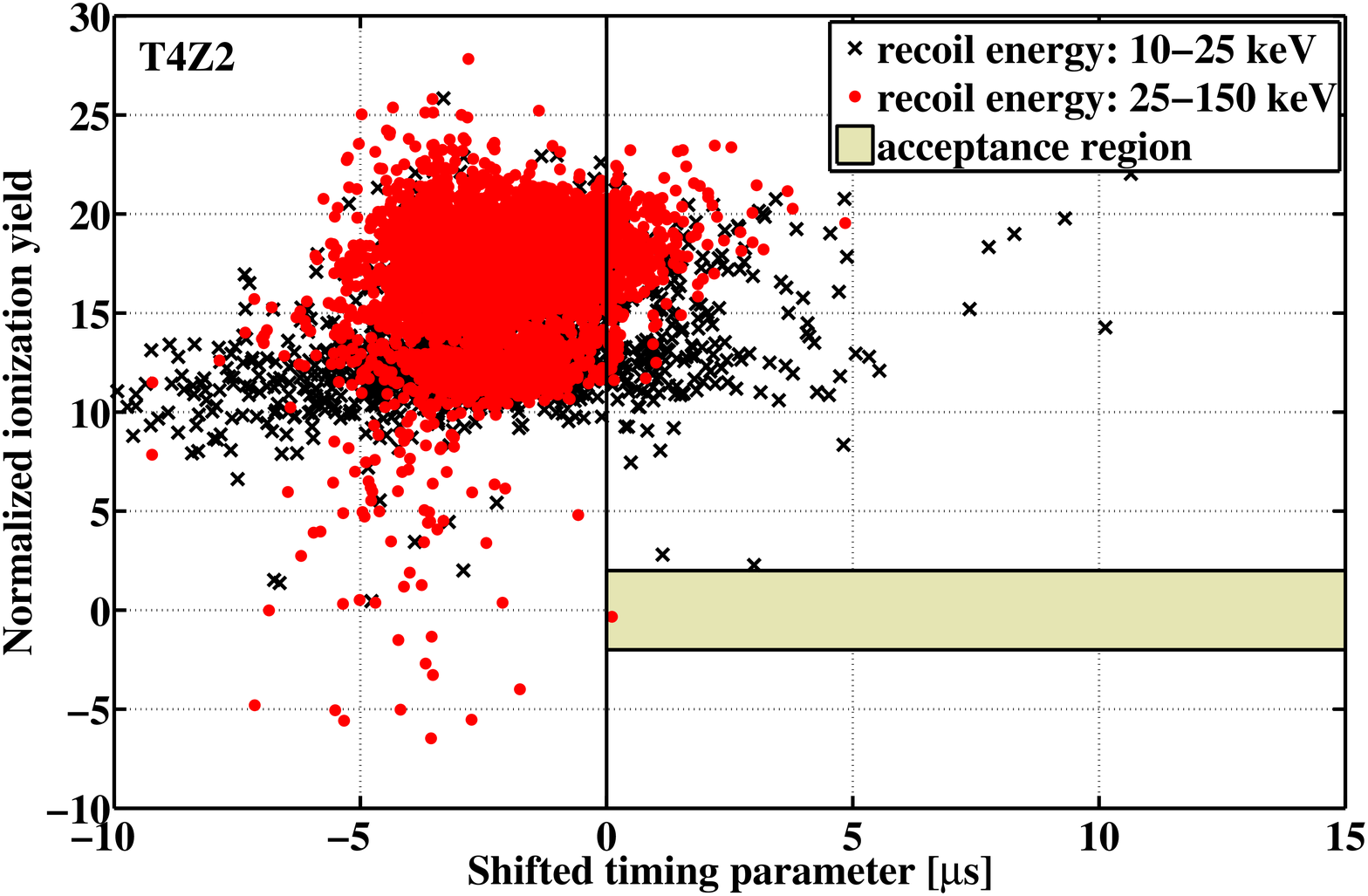}
\includegraphics[width=0.48\textwidth]{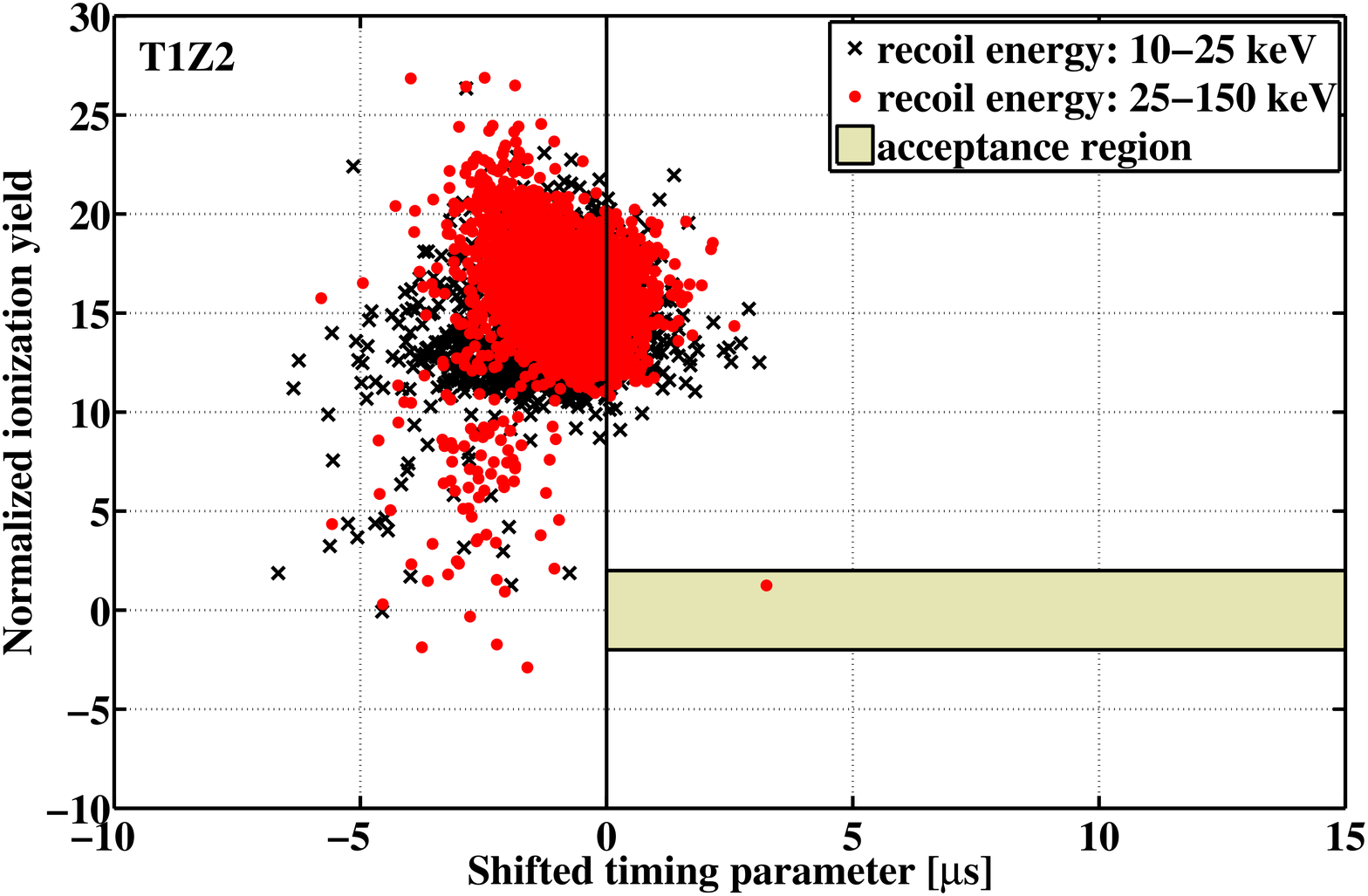}
\caption{(color online). Number of standard deviations each event is away from the mean of the nuclear-recoil band (normalized ionization yield) versus timing parameter relative to the timing-cut position (shifted timing parameter) for the three detectors with WIMP candidates above 25\,keV. The detectors are (from top to bottom) T4Z6, T4Z2 and T1Z2 with candidate events at 37.3\,keV,  73.3\,keV and  129.5\,keV, respectively. In detector T4Z6 three additional candidates occurred in the 10{\textendash}25\,keV range. The acceptance regions are indicated by the shaded boxes.}
\label{yieldtiming}
\end{figure}

After ``unblinding'', eleven events were observed within the acceptance region passing the surface-event rejection cut, three within the 25{\textendash}150\,keV range and eight between 10\,keV and 25\,keV. Figure 6 shows these candidates, along with all other WIMP-search events in or close to the signal region, which pass all constraints except for cuts on the ionization yield and timing parameter. As can be seen in Table~\ref{events}, the candidates are well distributed over the whole data-taking period and are spread over various detectors; though, six of the eleven candidates occurred in the two end cap detectors (T3Z6 and T4Z6), where there was less shielding from background, and where there was no detector below it to help reject background by detecting multiple scatters. It was verified that the performance of the experiment was stable at the times during which the events occurred. The detectors in which the three candidates above 25 keV occurred are examined in more detail in Fig.~\ref{yieldtiming}, where normalized ionization yield, defined as the distance from the nuclear-recoil band mean measured in units of standard deviations given by the width of the band, is plotted against the timing parameter relative to the actual cut position. 
\begin{table}[b!]
\begin{tabular}{cccc}
\toprule
Energy (keV) & Detector & Run & Date  \\
\hline
10.8 & T2Z3 & 127 & 31.05.2008 \\ 
11.8 & T4Z6 & 124 & 31.05.2007 \\
12.3* & T1Z5 & 125 & 27.10.2007 \\
12.8 & T3Z6 & 127 & 01.06.2008 \\
13.0 & T4Z6 & 125 & 05.10.2007 \\
14.7 & T3Z6 & 123 & 10.12.2006 \\
15.5* & T3Z4 & 125 & 05.08.2007 \\
16.4 & T4Z6 & 123 & 30.10.2006 \\
\hline
37.3 & T4Z6 & 126 & 02.02.2008 \\
73.3 & T4Z2 & 126 & 04.02.2008 \\
129.5 & T1Z2 & 123 & 24.12.2006\\
\botrule
\end{tabular}
\caption{Distribution of the eleven event candidates over detectors and time. The two marked events (*) are the candidates already found in our previous analysis of runs 125{\textendash}128 \cite{cdms2010}.}
\label{events}
\end{table}
The black/solid line denotes the timing-cut boundary on the given detector, and the shaded box indicates the acceptance region. The top plot is for T4Z6, with a WIMP candidate at 37.3\,keV and three additional candidates below 25\,keV. T4Z6 was a detector at the bottom of its tower with reduced background rejection capability. The middle plot shows events in T4Z2, where an event occurred just above the timing-cut boundary with a recoil energy of 73.3\,keV. Finally, we show events from T1Z2 in the bottom plot with a candidate above the analysis range from previous analyses at 129.5\,keV. This event is far above the timing-cut boundary and would be rejected neither by the surface-event cut from the previous analysis \cite{cdms2010}, nor by a tighter timing cut tuned to a leakage as low as 0.1 (instead of 0.6) events, which was the most stringent timing cut we tested. No additional events appear in the signal region above 25\,keV until the timing cut is loosened to an estimated surface-event leakage of more than 2.0 events.

The probability to observe three or more surface-leakage events between 25\,keV and 150\,keV given the background distribution $f(\mu)$ shown in Fig.~\ref{leakage} was calculated as
\begin{equation}
p = \int_0^{\infty} \textrm{d}\mu \, f(\mu) \cdot \sum_{k=3}^{\infty} \frac{e^{-\mu}\mu^k }{k!}
\end{equation}
and yields 9\%. 
Inclusion of the estimated neutron background increases this probability to 11\%, which is low but not negligible. Thus, this analysis does not constitute a significant detection of WIMP scattering. The probability for eight or more surface-background events between the 10\,keV threshold and 25\,keV was calculated based on a background distribution obtained analogously to the distribution in the 25{\textendash}150\,keV range and is 29\%, which indicates compatibility of our result with the background expectation. The inclusion of the neutron background has a negligible effect on the low-energy range due to the dominant surface-event background.

We used the optimum interval method \cite{optint} to compute the 90\%\,C.L. upper limit on the spin-independent cross section as a function of WIMP mass and splitting. All eleven WIMP candidates were included as possible signal, with no background subtraction. The differential rates were calculated under standard halo assumptions  according to \cite{savage}, which gives an updated version of the standard formula from \cite{lewinsmith}, correctly taking the effect of the Earth's velocity on the escape-velocity cutoff into account. We assumed this escape velocity $v_{\mathrm{esc}}$ to be 544\,km/s \cite{rave}, while the standard value of 220\,km/s was applied for the dispersion $v_0$ of the Maxwellian dark matter velocity distribution. Helm form factors and a three-dimensional parametrization of the Earth velocity $v_{\mathrm{E}}$ were used following \cite{lewinsmith}.

\begin{figure*}[t!]
\includegraphics[width=0.46\textwidth]{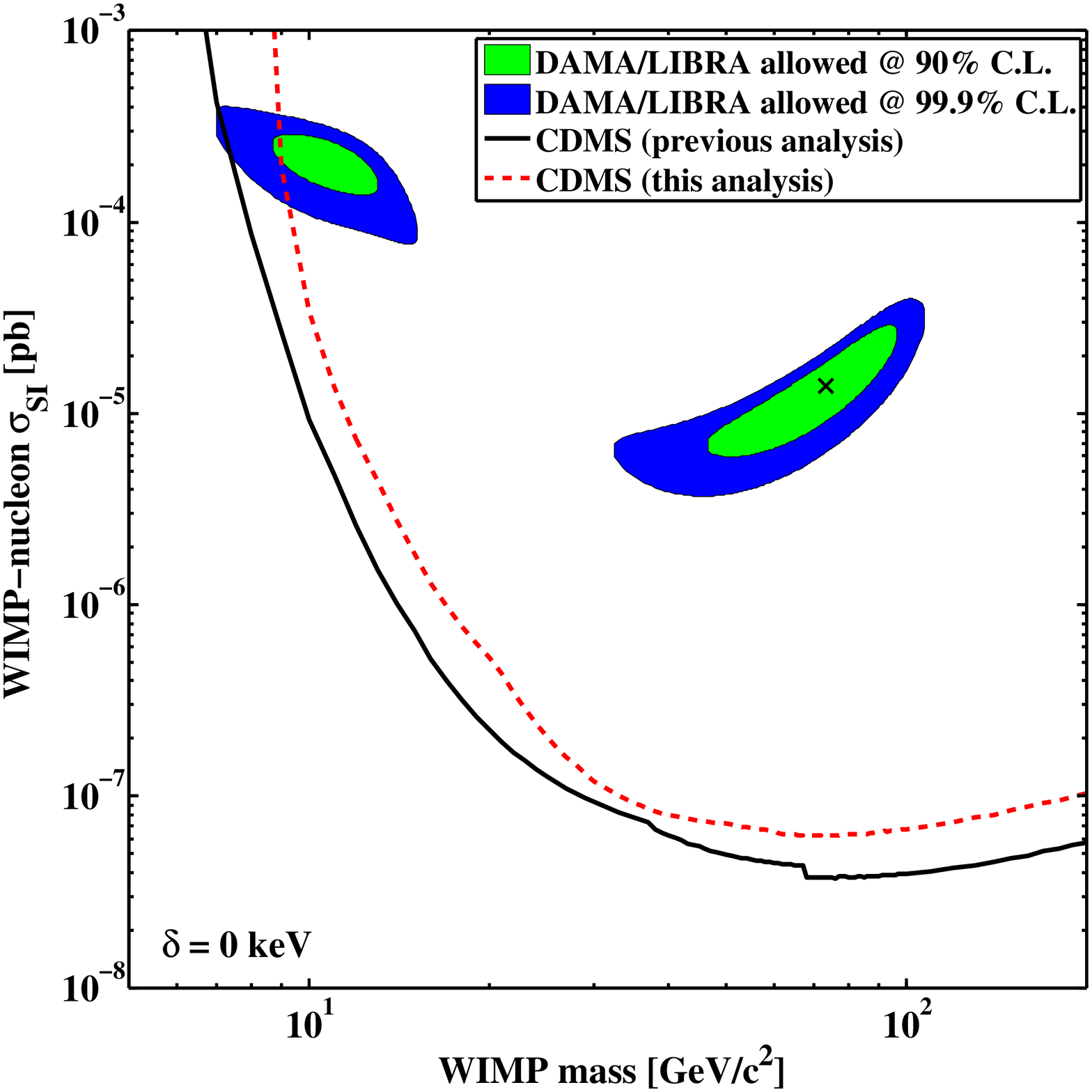}
\includegraphics[width=0.469\textwidth]{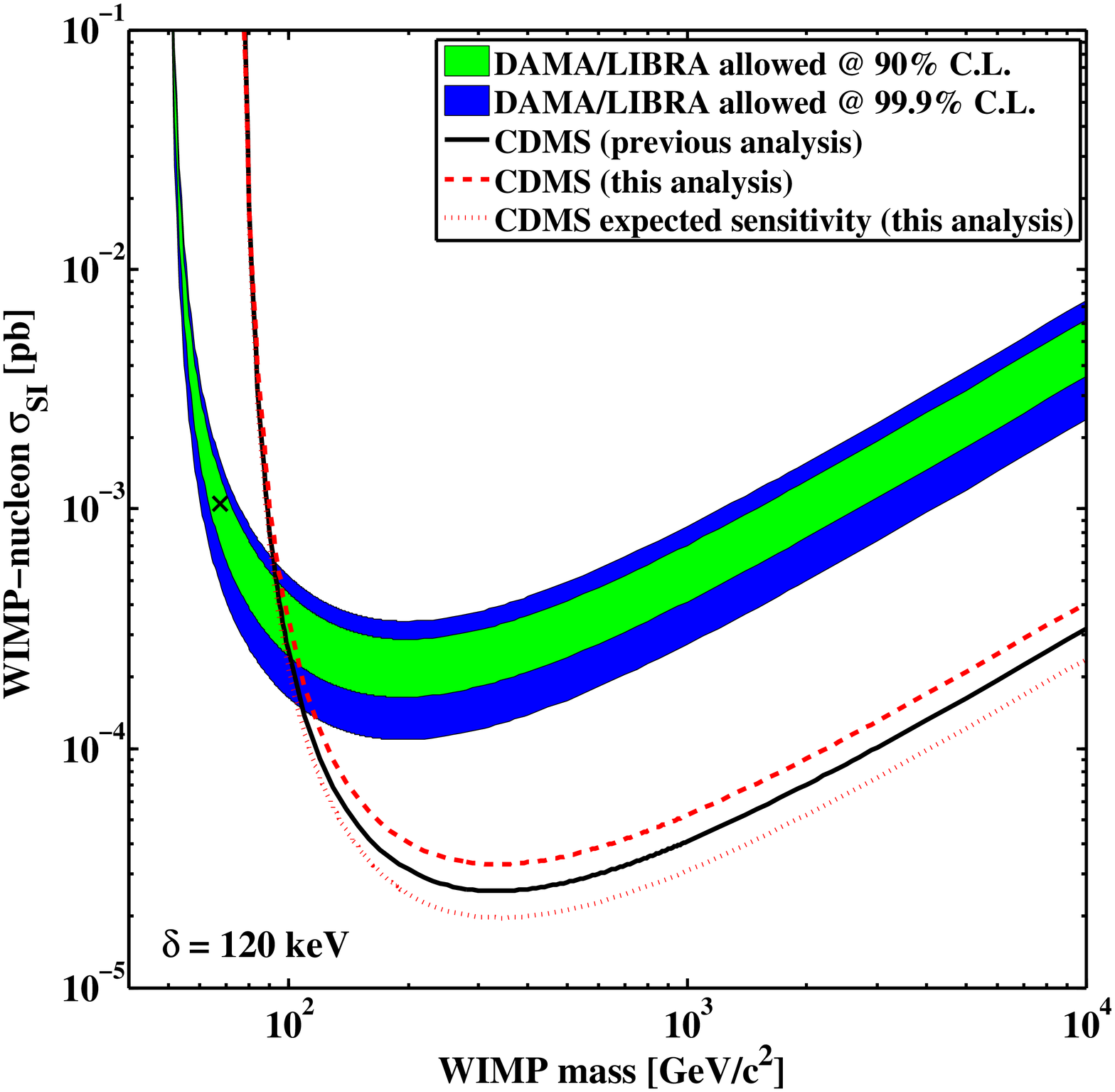}
\caption{(color online). 90\% C.L. upper limits on the scalar WIMP-nucleon cross section for WIMP-mass splittings of 0\,keV (left) and 120\,keV (right) from this analysis (red/dashed) and from our previous analysis (black/solid) \cite{cdms2010}. The red/dotted line in the right plot indicates the expected sensitivity for this analysis based on our estimate of the total background. The colored regions represent DAMA/LIBRA allowed regions at two different C.L.s (90,\,99.9\%) calculated following a \mbox{$\chi^2$ goodness-of-fit technique} \cite{savage}. The cross ($\times$) marks the parameter-space point which yields the minimum $\chi^2$ in the shown cross-section versus WIMP-mass plane given the WIMP-mass splitting.}
\label{crosssection}
\end{figure*}

Regions allowed by DAMA/LIBRA at two different C.L.s (90,\,99.9\%) were computed based on the published modulated spectrum in \cite{damalibra2010} from an exposure of 1.17\,ton-years. As in \cite{cdms2010}, we followed the $\chi^2$ goodness-of-fit technique advocated in \cite{savage} to investigate the compatibility between the results from DAMA/LIBRA and CDMS. Quenching factors of 0.30 and 0.09 were applied for Na and I nuclei in the DAMA/LIBRA setup, respectively \cite{damaquenching}. Possible channeling effects \cite{damachanneling} were not included in this study since they do not have a significant impact on the results from an iDM analysis where a signal is expected at tens of keV recoil energy \cite{channeling}.

Selected results from these computations are shown in Fig.~\ref{crosssection} in the cross-section versus WIMP-mass plane for two chosen WIMP-mass splittings. The left plot shows the standard case with $\delta = 0$\,keV, equivalent to assuming elastic scattering, while $\delta = 120$\,keV is used for the right plot, a value which is not experimentally excluded by our previous analysis. Apart from the DAMA/LIBRA allowed regions, and constraints emerging from the analysis presented in this paper, the plots also contain cross-section limits from our previous analysis of the 10{\textendash}100\,keV energy range \cite{cdms2010}. Constraints from the new analysis are less stringent. This was anticipated for the elastic scattering case and low WIMP-mass splittings in general, since more surface-background events were expected at low energies due to the looser timing cut defined in the 25{\textendash}150\,keV energy range. The limits are slightly weaker at $\delta = 120$\,keV, due to the occurrence of the three candidates above 25\,keV, where the rate is expected to peak for higher WIMP-mass splittings. The eight low-energy events have no effect on this part of the parameter space due to the utilization of the optimum interval method. WIMP masses above $\sim$100 GeV/c$^2$ are excluded for this mass splitting by the current and previous analysis.

\begin{figure}[b!]
\includegraphics[width=0.476\textwidth]{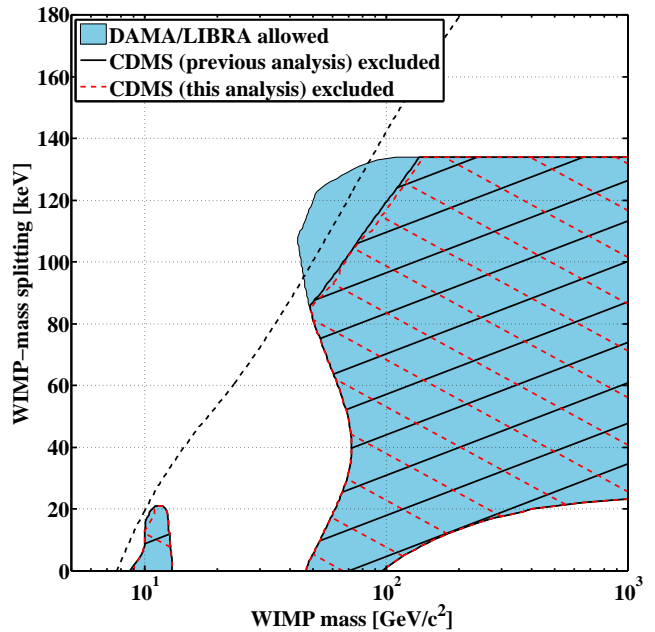}
\caption{(color online). The blue/shaded regions represent WIMP masses and WIMP-mass splittings for which cross sections exist that are compatible with the modulation spectrum observed by DAMA/LIBRA at 90\% C.L. The hatched regions show constraints on these parameters from the analysis presented in this paper (red/dashed) and from our previous analysis (black/solid) \cite{cdms2010}. The black/dashed line represents the maximum reach of the CDMS II experiment.}
\label{constraintsstandard}
\end{figure}
\begin{figure*}[t!]
\includegraphics[width=0.48\textwidth]{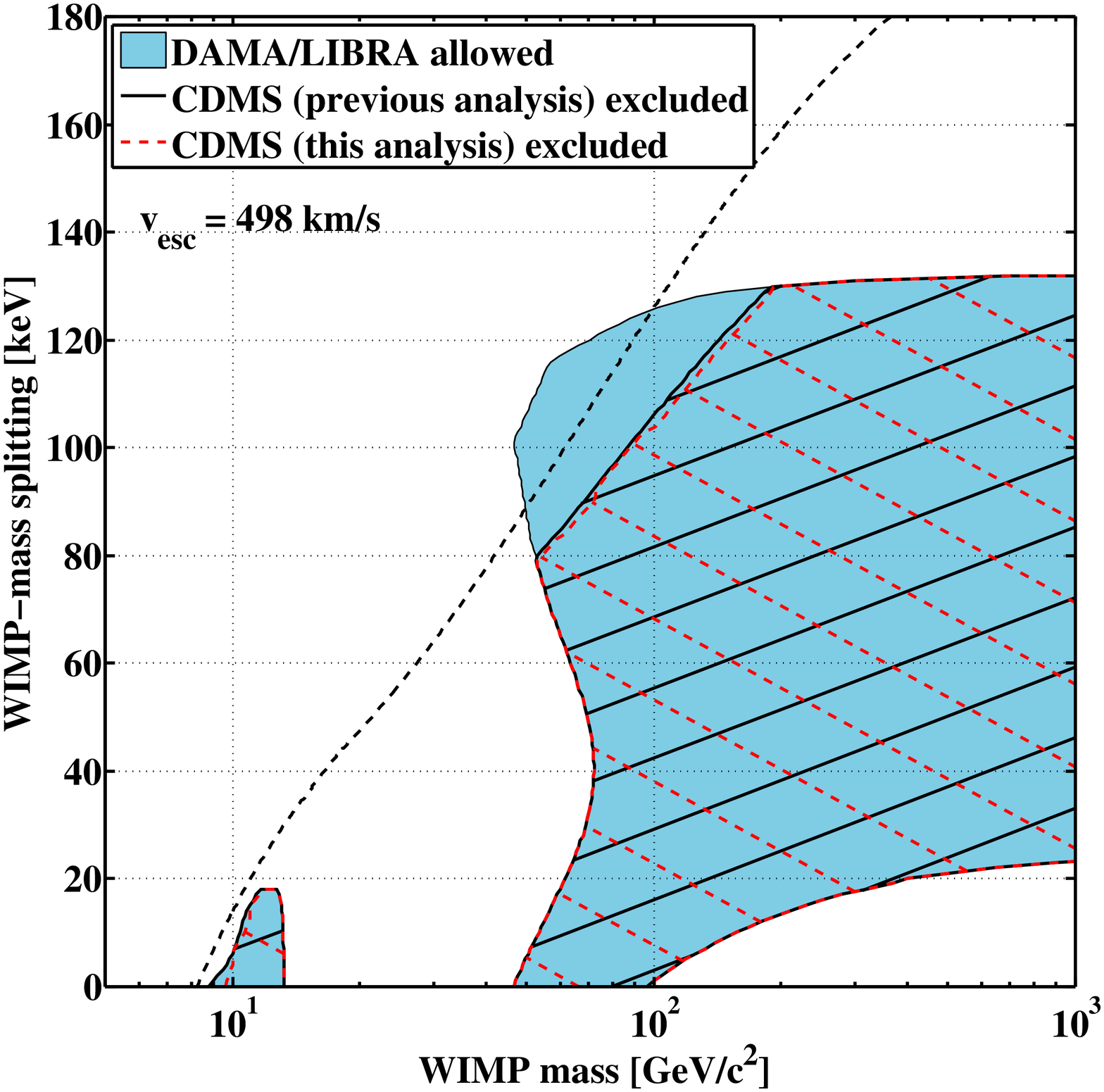}
\includegraphics[width=0.48\textwidth]{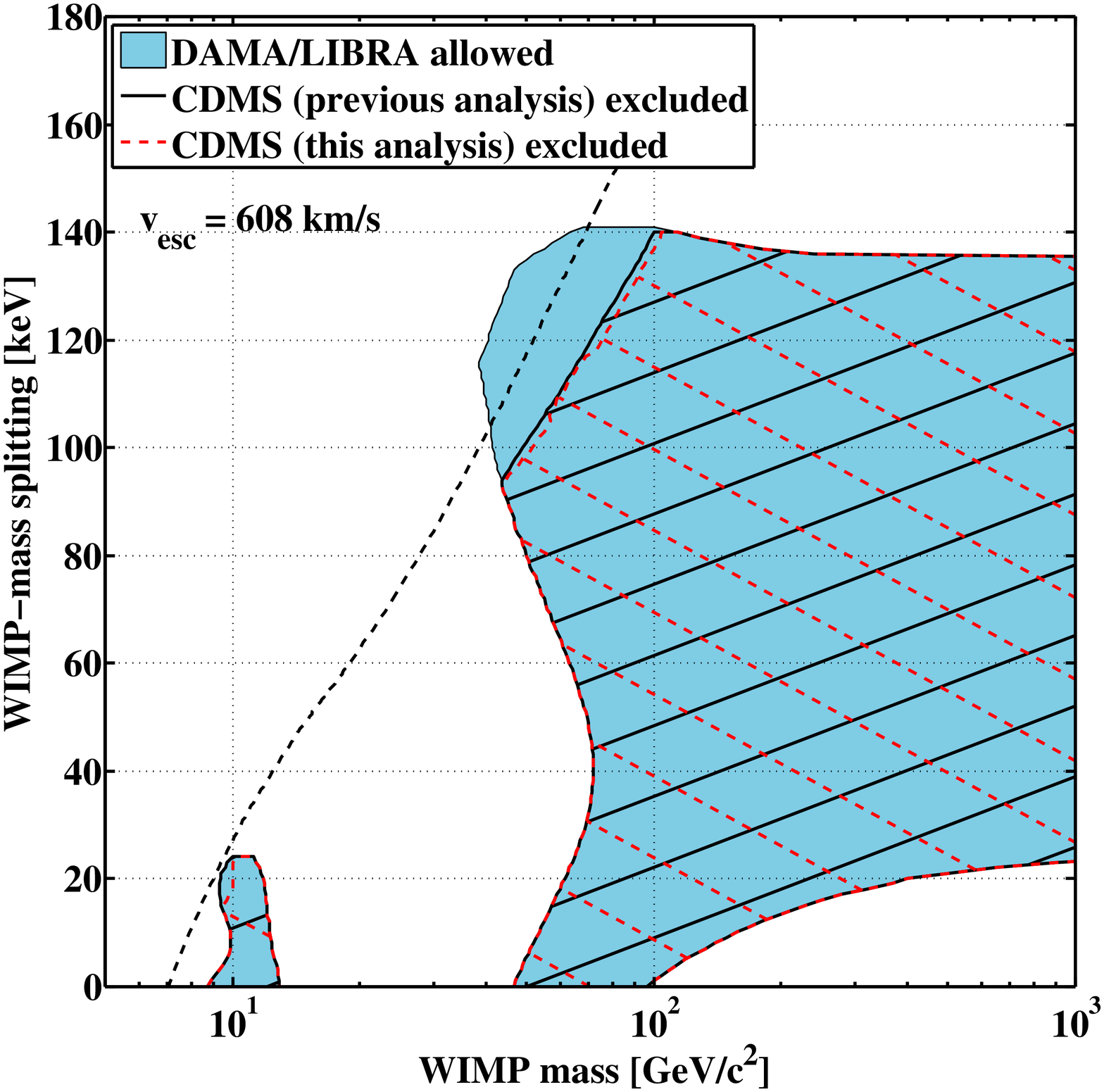}
\includegraphics[width=0.48\textwidth]{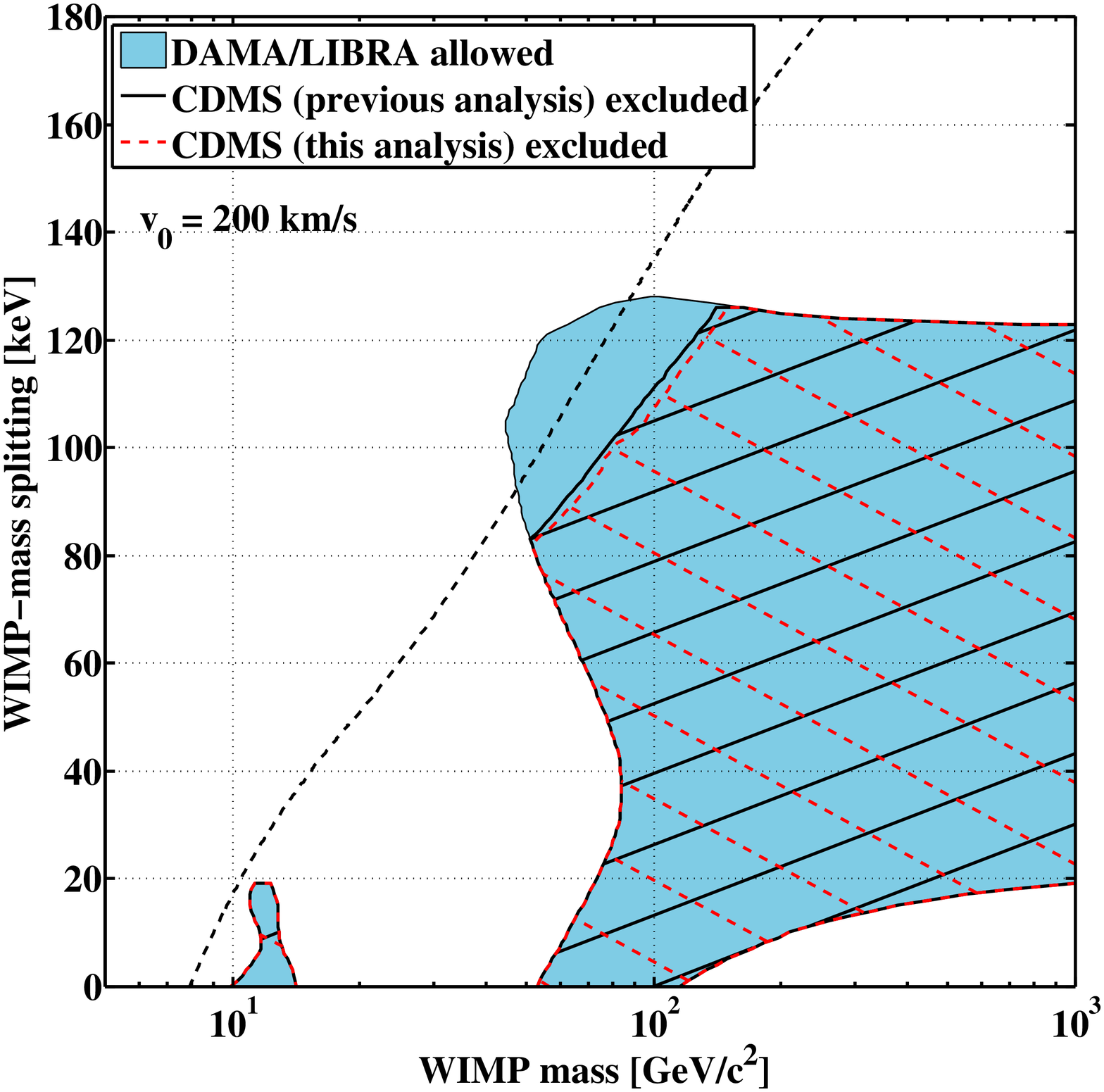}
\includegraphics[width=0.48\textwidth]{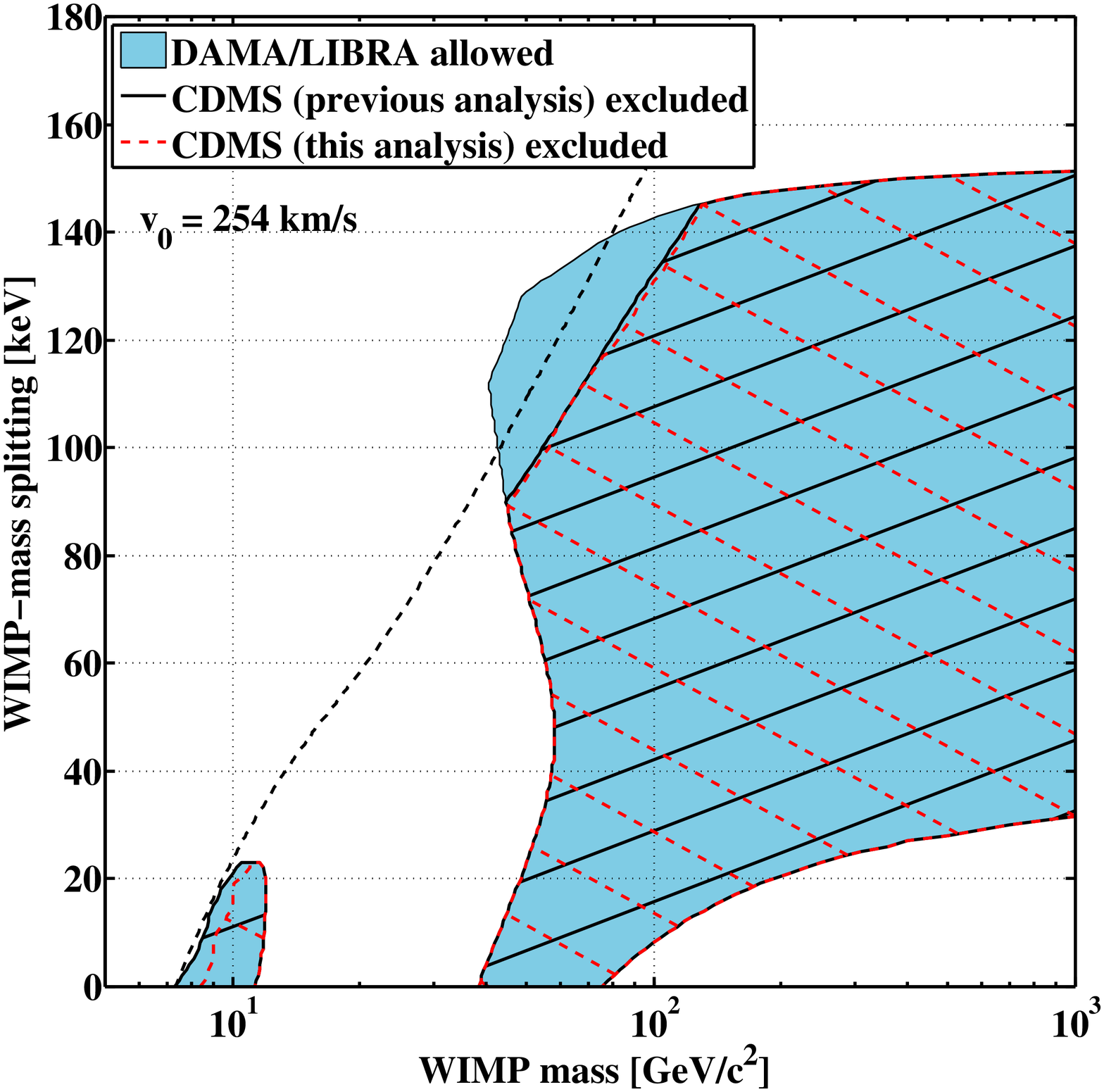}
\caption{(color online). Constraints from CDMS on the iDM parameter space allowed by DAMA/LIBRA. Same as Fig.~\ref{constraintsstandard} but with different velocity-distribution parameters, as given in the plots. All other parameters remain unchanged.}
\label{constraints}
\end{figure*}

Since the iDM parameter space is three-dimensional, consisting of the cross section, WIMP mass, and WIMP-mass splitting, we defined excluded regions by requiring the 90\% C.L. upper limit on the cross section from CDMS to completely rule out the corresponding DAMA/LIBRA allowed cross sections (also at 90\% C.L.) for given WIMP mass and WIMP-mass splitting. The results are shown in Fig.~\ref{constraintsstandard}. The only remaining allowed parameter space is within a narrow region at WIMP masses of \mbox{$\sim$100 GeV/c$^2$} and WIMP-mass splittings between \mbox{85\,keV and 135\,keV.} In the case of the new analysis presented in this paper there is also a tiny area in the low-mass region which is not excluded. The black/dashed line represents the maximum reach in the shown parameter space of an experiment using a Ge target like CDMS~II. It is computed based entirely on kinematics by demanding $v_{\mathrm{min}} = v_{\mathrm{esc}} + v_{\mathrm{E}}$, and is therefore independent of the cross-section parameter. Even with higher exposure and increased sensitivity, CDMS~II cannot rule out the entire DAMA/LIBRA allowed parameter space simply because (relative to an I nucleus) the Ge nucleus is too light. 
This is the main reason why the constraints from \mbox{ZEPLIN-III}  \cite{zeplin}, which employs a Xe target, are more stringent. Nevertheless, the results from CDMS~II currently have competitive sensitivity compared to the constraints from other Xe based experiments \cite{xenon10idm, zeplinII}.

The iDM scenario with a nonzero $\delta$ is particularly sensitive to the high-velocity tail of the dark matter velocity distribution due to the increased minimal velocity (see Eq.~\eqref{vmin}). Therefore, it exhibits a strong dependence on the velocity-distribution parameters $v_0$ (the dispersion) and $v_{\mathrm{esc}}$ (the galactic escape velocity) \cite{russell}. To examine these dependencies, in Fig.~\ref{constraints} we show plots similar to Fig.~\ref{constraintsstandard} but with different values of $v_{\mathrm{esc}}$ and $v_0$. The top plots explore the $v_{\mathrm{esc}}$ 90\% C.L. lower and upper limits found in \cite{rave} (498\,km/s and 608\,km/s), while all other parameters remain unchanged relative to Fig.~\ref{constraintsstandard}. In the bottom plots we varied only $v_0$, assigning a (convenient) lower value of 200\,km/s for the left plot and a higher value of 254\,km/s (the preferred value according to \cite{dispersion}) for the right plot. The capability of CDMS to constrain an iDM interpretation of the DAMA/LIBRA results is relatively independent of the actual velocity-distribution parameters. However, the shape and location of the parameter-space region, which is still allowed by CDMS, has a considerable dependence on $v_{\mathrm{esc}}$ and $v_0$, as expected. Non-Maxwellian velocity distributions as discussed in \cite{simulations, nonmaxwell} are beyond the scope of this study.

In this paper we presented the first CDMS analysis which includes recoil energies up to 150 keV. The entire five-tower data set was used in a combined analysis. Because of the occurrence of the three candidate events between 25\,keV and 150\,keV the constraints on the iDM parameter space are slightly weaker than from our previous analysis for which no events were observed at intermediate energies where the rate is expected to peak. The only remaining parameter space allowed by CDMS data is within a narrow region at WIMP masses of \mbox{$\sim$100 GeV/c$^2$} and WIMP-mass splittings between \mbox{85\,keV and 135\,keV}, assuming standard values for the WIMP-velocity distribution parameters. Varying the values of these parameters changes this region considerably but has only a minor effect on the capability of CDMS to constrain an iDM interpretation of the DAMA/LIBRA results. Finally, though this analysis was performed with regard to the iDM scenario, the expansion of the analysis range to 150\,keV could be useful to test other models predicting a signal at tens of keV recoil energy.

The CDMS collaboration gratefully acknowledges the contributions of numerous engineers and technicians; we would like to especially thank Jim Beaty, Bruce Hines, Larry Novak, Richard Schmitt and Astrid Tomada.  In addition, we gratefully acknowledge assistance from the staff of the Soudan Underground Laboratory and the Minnesota Department of Natural Resources. This work is supported in part by the 
National Science Foundation (Grant Nos.\ AST-9978911, PHY-0542066, 
PHY-0503729, PHY-0503629,  PHY-0503641, PHY-0504224, PHY-0705052, PHY-0801708, PHY-0801712, PHY-0802575 and PHY-0855525), by
the Department of Energy (Contract Nos. DE-AC03-76SF00098, DE-FG02-91ER40688, 
DE-FG02-92ER40701, DE-FG03-90ER40569, and DE-FG03-91ER40618), by the Swiss National 
Foundation (SNF Grant No. 20-118119) and by NSERC Canada (Grant SAPIN 341314-07).

\end{document}